\documentclass[12pt]{amsart}

\newcommand{\R}{\mathbb R}

\newcommand{\Symb}{Symb}
\newcommand{\Diff}{Diff}

\newcommand{\reg}{\text{reg}}
\newcommand{\x}{\mathbf x}
\newcommand{\y}{\mathbf y}

\begin{document}

\title[Differential operators and quantum field theory]
{Differential operators on infinite dimensional space and quantum field theory}
\author{A. V. Stoyanovsky}
\thanks{Partially supported by the grant RFBR 10-01-00536.}
\email{alexander.stoyanovsky@gmail.com}
\address{Russian State University of Humanities}

\begin{abstract}
We conjecture that the renormalized perturbative $S$-matrix of quantum field theory coincides with the evolution operator
of the standard functional differential Schrodinger equation whose right hand side (quantum local Hamiltonian)
is understood as an element of an appropriate quantization of the Poisson algebra of classical field theory Hamiltonians.
We show how to construct a quantization of this algebra, close to the algebra of differential operators on infinite
dimensional space, but seemingly not appropriate for quantum field theory.
\end{abstract}

\maketitle

\section{Introduction: the Schrodinger FDE}

Giving mathematical sense to quantum field theory is one of the main fundamental open problems of natural science.
The achievements of constructive field theory in low dimensions lead us to believe that a solution of this problem
indeed does exist. In the present paper we outline a program of possible reduction of this problem to a problem of
theory of differential equations, mathematically similar to usual quantum mechanics.
The differential equations that we consider are not usual PDE's but
rather linear differential equations on infinite dimensional functional space called {\it functional
\emph(or variational\emph)
differential equations} (FDEs). The main equation under our consideration is the so called {\it Schrodinger FDE}
\begin{equation}
ih\frac{\partial\Psi}{\partial t}=\widehat H(t)\Psi,
\end{equation}
where $\Psi$ is a (not well defined yet) {\it Schrodinger wave functional}, and
\begin{equation}
\begin{aligned}{}
\hat H(t)&=\int\frac12\left(\hat\pi(\x)^2+\sum_{j=1}^n\left(\frac{\partial\hat\varphi}{\partial x_j}\right)^2(\x)
+m^2\hat\varphi(\x)^2\right)d\x\\
&+\int\left(\frac1{k!}g(t,\x)\hat\varphi(\x)^k+j(t,\x)\hat\varphi(\x)\right)d\x,\\
\end{aligned}
\end{equation}
where $\x=(x_1,\ldots, x_n)$; $\hat\varphi(\x)=\varphi(\x)$
and $\hat\pi(\x)=-ih\frac{\delta}{\delta\varphi(\x)}$ are the operators satisfying the canonical commutation relations
\begin{equation}
[\hat\varphi(\x),\hat\pi(\x')]=ih\delta(\x-\x'),\ \ [\hat\varphi(\x),\hat\varphi(\x')]=[\hat\pi(\x),\hat\pi(\x')]=0;
\end{equation}
$k$ is a natural number; $m$ is a real number called mass;
and $g(t,\x)$, $j(t,\x)$ are scalar functions called respectively the interaction cutoff
function and the source. For simplicity, in this paper we consider only the
Schrodinger FDE for scalar field with self-act\-ion and for distinguished time axis,
generalization to the vector case and to the relativistically invariant case [1,2] being straightforward. Equation (1)
has been written out non-de\-numer\-able number of times in the literature, but, to the best of our knowledge, it has been
usually considered purely formally (except for several papers by I.~Kanatchikov [3,4] and the author [1,2]).

If we try to consider equation (1) mathematically, then we immediately arrive at the problem of developing
mathematical calculus
of differential operators like (2). For example, it is unclear already how to consider $\widehat H(t)^2$. Definitely, it
should not be a usual functional differential operator [5] in which all the $\hat\pi$'s stand to the right of
$\hat\varphi$'s. It also does not belong to infinite dimensional Weyl--Moyal algebra (even for $m=1$, $j\equiv g\equiv0$).

Our approach to the algebra of quantum Hamiltonians like (2) is that it should be a {\it quantum deformation}
of the Poisson algebra of classical field theory Hamiltonians (also called symbols). We denote this Poisson algebra by
$\Symb$. The definition of $\Symb$ and a motivation for it have been given in [6,7]. We recall this definition
in \S2. After that, in \S3, we construct a quantization of the algebra $\Symb$ which is a natural candidate
for the algebra of differential operators on the Schwartz space. Finally, in \S4 we present the conjecture on the
$S$-matrix.

I would like to thank I. V. Kanatchikov and V. P. Maslov for helpful discussions. I am also grateful to Professor
Alberto S. Cattaneo for bringing the paper [11] to my attention.

\section{The Poisson algebra of classical field theory Hamiltonians [6,7]}

Denote by $S$ the Schwartz space of smooth real functions on $\R^n$ rapidly decreasing at infinity, so that
$\varphi,\pi\in S$.

{\bf Definition.} A continuous polynomial functional
$H(\varphi,\pi)$ on $S\times S$ is called a Hamiltonian (or a symbol) if its first
functional differential $\delta H$, which is a linear functional
on test functions $(\delta\varphi,\delta\pi)\in S\oplus S$ for every $(\varphi,\pi)$, belongs to $S\oplus S\subset
S'\oplus S'$ (here $S'$ is the space of tempered distributions dual to $S$). Moreover, $\delta H$ should be
infinitely differentiable as an $S\oplus S$-valued functional on $S\times S$.

Denote the space of symbols by $\Symb$.
\medskip

{\bf Proposition.} $\Symb$ is a topological Poisson algebra with respect to the
standard Poisson bracket
\begin{equation}
\{H_1,H_2\}=\int\left(\frac{\delta H_1}{\delta\pi(\x)}\frac{\delta H_2}{\delta\varphi(\x)}
-\frac{\delta H_1}{\delta\varphi(\x)}\frac{\delta H_2}{\delta\pi(\x)}\right)d\x.
\end{equation}
\medskip

{\it Proof} is straightforward.
\medskip

Note that, unlike any differentiable functional on the space $S\times S$,
any $H\in\Symb$ generates a well defined Hamiltonian flow on the phase space $S\times S$.

Any $H\in\Symb$ has the form
\begin{equation}
\begin{aligned}{}
H(\varphi,\pi)&=\sum_{k=0}^N\sum_{l=0}^M H_{k,l}(\varphi,\pi),\\
H_{k,l}(\varphi,\pi)&=\frac1{k!l!}\int a_{k,l}(\x_1,\ldots,\x_k;\y_1,\ldots,\y_l)\\
&\times\varphi(\x_1)\ldots\varphi(\x_k)\pi(\y_1)\ldots\pi(\y_l)d\x_1\ldots d\x_k d\y_1\ldots d\y_l
\end{aligned}
\end{equation}
for certain tempered distributions $a_{k,l}$ symmetric in $\x_1,\ldots,\x_k$ and in $\y_1,\ldots,\y_l$ (by the Schwartz
kernel theorem).
\medskip

{\bf Definition.} If all $a_{k,l}$ are smooth functions rapidly decreasing at infinity then the Hamiltonian
$H$ is called {\it regular}. Otherwise it is called {\it singular}.
\medskip

Denote the subspace of $\Symb$ consisting of regular Hamiltonians by $\Symb^{\reg}$. Clearly,
it is a Poisson subalgebra dense in $\Symb$.

\section{A quantization of the algebra $\Symb$ and renormalization on a space-like surface}

\subsection{Quantization of $\Symb$}
We shall not recall here the well-known definition of quantization of the Poisson algebra $\Symb$. For two slightly
different exact definitions, see [6,7]. Here we present a solution to the existence problem for quantizations of
$\Symb$. Namely, we construct a filtered topological associative algebra which is a natural candidate for the role of the
algebra of differential operators on the space $S$, whose associated graded algebra coincides with $\Symb$.
This construction is close to the construction of Lie algebroids (see, for example, [8]).

{\bf Definition.} A symbol
\begin{equation}
\partial=v(\varphi,\pi)+f(\varphi)
\end{equation}
is called a symbol of first order if $f,v\in\Symb$ and $v$ is linear in $\pi\in S$.

The symbols of first order form a complex topological Lie algebra with respect to the Poisson bracket (4) multiplied by
$h$. Denote its
topological universal enveloping algebra (the quotient of the topological tensor algebra by the usual ideal)
by $\widetilde{\Diff}_h$. Denote the product in this algebra by $*$.
Further, denote by $\Diff=\Diff_h$ the topological quotient of this algebra by the closure of the two-sided ideal
generated by the relations
\begin{equation}
\begin{aligned}{}
v*f&=fv+\frac h2\{v,f\},\\
f_1*f_2&=f_1f_2.
\end{aligned}
\end{equation}
The coefficient $\frac h2$ is chosen so that the algebra $\Diff$ has a natural antilinear involution
$\partial\to\bar\partial$ such that for real $f,v$ we have $\bar f=f$, $\bar v=-v$, and
\begin{equation}
\overline{\partial_1*\partial_2}=\bar\partial_2*\bar\partial_1.
\end{equation}

An advantage of this construction is that it seemingly gives a quantization of the algebra $\Symb$. Its disadvantage
is that in the algebra $\Diff$, it is difficult to compute with explicit expressions like (2).

\subsection{Renormalization on space-like surface [7]} Clearly, we have a usual inclusion
\begin{equation}
\Symb^\reg\hookrightarrow\Diff
\end{equation}
where in $\Symb^\reg$ all the $\hat\pi$'s are put to the right of $\hat\varphi$'s. The corresponding product in
$\Symb^\reg$ reads [7]
\begin{equation}
\begin{aligned}{}
H_1*_{Diff}H_2(\varphi,\pi)=&\exp\left(ih\int\frac{\delta}{\delta\pi_1(\x)}\frac{\delta}{\delta\varphi_2(\x)}d\x\right)\\
&H_1(\varphi_1,\pi_1)H_2(\varphi_2,\pi_2)|_{\varphi_1=\varphi_2=\varphi,\pi_1=\pi_2=\pi}.
\end{aligned}
\end{equation}
On the other hand, if we want to multiply local expressions like (2), then we should
(non-canonically) identify $\Diff$ with $\Symb$, and the restriction of multiplication $*$
to $\Symb^\reg$ will give another multiplication on $\Symb^\reg$. The difference between the two multiplications can be
measured by a complex linear map
\begin{equation}
R=Id+O(h):\Symb^\reg\to\Symb^\reg
\end{equation}
called {\it renormalization on a space-like surface}.

In other words, the multiplication $*$ on $\Diff$ is not well suited to passing to regularization of a Hamiltonian.

\section{The main conjecture}

{\bf Conjecture.} For certain quantization of $\Symb$, for $n=3$ and for $k=4$ in (1), the renormalized
perturbative Bogoliubov $S$-matrix [10] coincides with the perturbative expansion of the evolution operator of the
Schrodinger FDE (1).

The arguments for this Conjecture are essentially the same as that in [9], \S2. The difference with [9] is that in that
paper we did not take into account renormalization on space-like surfaces.

We expect that the quantization of the algebra $\Symb$ needed for the Conjecture is not the one constructed in \S3.1
above.
One should rather use in it creation and annihilation operators $\varphi_+$ and $\varphi_-$ instead of $\varphi$ and $\pi$,
cf. [11].


\begin{thebibliography}{99}
\bibitem{1} A. V. Stoyanovsky, Introduction to the mathematical principles of quantum field theory
(in Russian), Moscow, Editorial URSS, 2007.
\bibitem{2} A. V. Stoyanovsky, Generalized Schrodinger equation for free field, hep-th/0601080.
\bibitem{3} I. V. Kanatchikov, Precanonical quantization and the Schrodinger wave functional, hep-th/0012084.
\bibitem{4} I. V. Kanatchikov, Precanonical quantization and the Schrodinger wave functional revisited, arXiv: 1112.5801
[hep-th].
\bibitem{5} V. G. Zadorozhnii, Methods of variational analysis, Regular and Chaotic Dynamics, Institute for Computer
Research, Moscow, 2006 (in Russian).
\bibitem{6} A. V. Stoyanovsky,
The Poisson algebra of classical Hamiltonians in field theory and the problem of its quantization,
arXiv:1008.3333 [math-ph].
\bibitem{7} A. V. Stoyanovsky, On the mathematical sense of renormalization, arXiv:1110.0002 [physics.gen-ph].
\bibitem{8} A. A. Beilinson, V. G. Drinfeld, Chiral algebras, Colloquium publications, vol. 51, AMS, Providence, 2004.
\bibitem{9} A. V. Stoyanovsky, No-Counterterm approach to quantum field theory, arXiv:1002.0915 [hep-th].
\bibitem{10} N. N. Bogoliubov and D. V. Shirkov, Introduction to the theory of quantized fields, John Wiley and Sons,
1976, 3rd edition.
\bibitem{11} M. D\"utch and K. Fredenhagen, Perturbative algebraic field theory, and deformation quantization,
hep-th/0101079.
\end{thebibliography}
\end{document}